\documentclass[aps,prb,superscriptaddress,twocolumn,amsmath,showpacs,amssymb]{revtex4}
\usepackage{amssymb}
\usepackage[dvips]{graphicx}
\usepackage{dcolumn}
\usepackage{bm}
\usepackage{epsfig}
\usepackage[latin1]{inputenc}

\begin{document}

\title{Physical properties and magnetic structure of TbRhIn$_{5}$ intermetallic compound.}

\author{R. Lora-Serrano}
\affiliation{Instituto de F\'isica "Gleb Wataghin",UNICAMP,13083-970, Campinas-SP, Brazil.}
\email{rlora@ifi.unicamp.br}

\author{C. Giles}
\affiliation{Instituto de F\'isica "Gleb Wataghin",UNICAMP,13083-970, Campinas-SP, Brazil.}

\author{E. Granado}
\affiliation{Instituto de F\'isica "Gleb Wataghin",UNICAMP,13083-970, Campinas-SP, Brazil.}
\affiliation{Laborat\'orio Nacional de Luz S\'incrotron, Caixa Postal 6192, CEP 13084-971 Campinas-SP, Brazil}

\author{D. J. Garcia}
\affiliation{Instituto de F\'isica "Gleb Wataghin",UNICAMP,13083-970, Campinas-SP, Brazil.}

\author{E. Miranda}
\affiliation{Instituto de F\'isica "Gleb Wataghin",UNICAMP,13083-970, Campinas-SP, Brazil.}

\author{O. Ag\"{u}ero}
\affiliation{Instituto de F\'isica "Gleb Wataghin",UNICAMP,13083-970, Campinas-SP, Brazil.}

\author{L. Mendon\c{c}a Ferreira}
\affiliation{Instituto de F\'isica "Gleb Wataghin",UNICAMP,13083-970, Campinas-SP, Brazil.}

\author{J. G. S. Duque}
\affiliation{Instituto de F\'isica "Gleb Wataghin",UNICAMP,13083-970, Campinas-SP, Brazil.}

\author{P. G. Pagliuso}
\affiliation{Instituto de F\'isica "Gleb Wataghin",UNICAMP,13083-970, Campinas-SP, Brazil.}

\date{\today}

\begin{abstract}

In this work we report the physical properties of the new intermetallic compound TbRhIn$_{5}$ investigated by
means of temperature dependent magnetic susceptibility, electrical resistivity, heat-capacity and resonant x-ray magnetic 
diffraction experiments. TbRhIn$_{5}$ is an intermetallic compound that orders antiferromagnetically at $T_N$
= 45.5 K, the highest ordering temperature among the existing RRhIn$_{5}$ (1-1-5, R = rare earth) materials.
This result is in contrast to what is expected from a de Gennes scaling along the RRhIn$_{5}$ series. The X-ray
resonant diffraction data below $T_N$ reveal a commensurate antiferromagnetic (AFM) structure with a propagation
vector ($\frac{1}{2}$ 0 $\frac{1}{2}$) and the Tb moments oriented along the \textit{c}-axis. Strong (over two order of
magnitude) dipolar enhancements of the magnetic Bragg peaks were observed at both Tb absorption edges $L_{II}$ and $L_{III}$, indicating a fairly high polarization of the Tb 5\textit{d} levels. Using a mean field model including an isotropic first-neighbors
exchange interaction (J$_{R-R}$) and the tetragonal crystalline electrical field (CEF), we were able to fit our
experimental data and to explain the direction of the ordered Tb-moments and the enhanced $T_N$ of this
compound. The evolution of the magnetic properties along the RRhIn$_{5}$ series and its relation to CEF effects
for a given rare-earth is discussed.

\end{abstract}

\pacs{75.50.Ee, 75.30.Gw, 75.10.Dg, 75.20.En}

\maketitle


\section{\bf INTRODUCTION}

The discovery of new series of structurally related compounds with novel physical behavior is an important
approach to explore fundamental problems in the physics of highly correlated electron systems. Such problems
include the interplay between superconductivity, magnetism and heavy fermion behavior in structurally related
materials.\cite{Fisk1,Moriya} Although the microscopic origin of this interplay remains a mystery, the search
for new heavy fermion superconductors (HFS) is partially guided by the knowledge that certain structures favor
the formation of this heavy electron ground state. Remarkable examples of amazing physical properties occurring
in structurally related compounds are the series Ce$_{m}$MIn$_{3m+2}$ (M = Co, Rh or Ir, m =
1,2)\cite{Hegger,Petrovic1,Petrovic2,Thompson,Chen} and their Pu-based analogs
PuMGa$_{5}$.\cite{Sarrao,eric,Curro} These materials grow in a tetragonal variant of the Cu$_{3}$Au-type
structure and can be viewed as layers of a cubic cell (CeIn$_{3}$ or PuGa$_{3}$) stacked sequentially along the
\textit{c}-axis with intervening layers of M(In,Ga)$_{2}$.\cite{Moshopoulou} The discovery of unconventional
superconductivity (USC) in many of these compounds is an exciting opportunity to further explore the possibility
of magnetically mediated superconductivity in strongly correlated electrons systems and its relationship with
dimensionality and crystal structures. For instance, systematic alloying studies in
CeRh$_{1-x}$Ir$_{x}$In$_{5}$\cite{pagliuso1,pagliuso2} and PuCo$_{1-x}$Rh$_{x}$Ga$_{5}$\cite{eric} (PuCoGa$_{5}$
possesses the highest superconducting transition temperature, $T_c$ = 18 K, among the pure Pu-based
compounds\cite{Sarrao}) have revealed an intriguing linear dependence between $T_c$ and the ratio of the
tetragonal lattice parameters \textit{c}/\textit{a} at ambient pressure, indicating that the increasing of the quasi-2D
character of their crystal structure may favor USC.

Such as the high-$T_c$ and the organic superconductors, HFS are believed to be magnetically mediated
superconductors.\cite{Mathur} Furthermore, in most cases, USC seems to occur at the vicinity of a magnetically
ordered state and the spin fluctuations (SF) associated with this magnetic phase may mediate the superconducting
pair formation.\cite{Mathur} More recently, a bridge connecting the HFS and high-$T_c$ superconductors (HTSC) has
been proposed by NMR studies in PuCoGa$_{5}$. These studies have shown that superconductivity in this material
is unconventional with d-wave symmetry and similar properties to those found in other HFS and HTSC, suggesting
that these classes of complex USC may share the same pairing mechanism.\cite{Curro}

Focusing our attention now to the Ce-based HFS, the magnetic properties of these materials are associated with
their 4\textit{f} electrons. Although these compounds display obvious heavy-fermion behavior, evidences for
low-temperature 4\textit{f} local moment behavior has also been found in this family. \cite{Takeuchi,Alver,Victor} This
apparent contradiction is directly related to an open question in condensed matter physics regarding the details
of the crossover between quasi-localized 4\textit{f} electron magnetic behavior at high-T to a renormalized
heavy-electron state at lower temperatures. However, if certain structures favor USC mediated by magnetic
fluctuations, it is an important first step to understand how layered structures can affect CEF anisotropy,
magnetic exchange and/or anisotropic transport properties (quasi-2D band-structure). In this regard, detailed
studies of the 4\textit{f} electrons magnetism along the series of rare-earth and actinides based 1-1-5 compounds may
be very elucidative.

In the case of the Ce-based compounds, their Nd-, Sm- and Gd-based structurally related magnetic materials have
been studied in detail.\cite{pagliuso3,pagliuso4,granado1,granado2,wei,wei1,wei2} It has been found that the
magnetic properties of these non-Ce analogs mainly depend on the interplay between CEF effects and exchange
magnetic interaction. For example, among the Nd-based compounds NdMIn$_5$ and Nd$_{2}$MIn$_{8}$ analogs for M =
Rh or Ir, it was found a systematic relationship between the AFM ordering temperature $T_N$ and the low-\textit{T} CEF
splitting.\cite{pagliuso4} Besides, when the magnetic properties of the tetragonal variants NdMIn$_{5}$ and
Nd$_{2}$MIn$_{8}$ are compared to that for its cubic NdIn$_{3}$, $T_N$ is enhanced by a factor of
two.\cite{pagliuso2,pagliuso4,raimundo}In contrast, the tetragonal CeRhIn$_{5}$ and Ce$_{2}$RhIn$_{8}$ present a
$T_N$ a factor of two smaller than that for CeIn$_{3}$ whereas for the Gd-based tetragonal materials the low
temperature magnetic properties remain nearly unaltered compared to the
GdIn$_{3}$.\cite{pagliuso2,pagliuso4,granado2}Finally, the resolved magnetic structures for some of these
compounds have revealed that the rare-earth magnetic moments lie in the \textit{ab}-plane for CeRhIn$_{5}$ and
GdRhIn$_{5}$ compounds and point along the \textit{c}-axis for NdRhIn$_{5}$ tetragonal
materials.\cite{wei,wei1,wei2,granado2}

To further explore the trends in the evolution of the magnetic properties within the series, particularly for the
RRhIn$_{5}$ compounds, we have investigated the physical properties and magnetic structure of a new member of
this series, TbRhIn$_{5}$. It was found to be an intermetallic compound that orders antiferromagnetically at
$T_N$ $\sim$ 46 K, the highest ordering temperature among the existing RRhIn$_{5}$ materials. The magnetic
structure of TbRhIn$_{5}$, resolved using Resonant X-ray Magnetic Diffraction (RXMD) experiments, is commensurate with
propagation vector $\tau$=($\frac{1}{2}$ 0 $\frac{1}{2}$) and the Tb moments oriented along the \textit{c}-axis. The direction of the ordering Tb-moments and the enhanced $T_N$ of this
compound were successfully explained by a mean field model including an isotropic first-neighbors exchange
interaction (J$_{R-R}$) and the tetragonal CEF, and was used to fit the magnetic susceptibility and specific heat experimental data.

Besides revealing the magnetic properties of this novel compound, the reported results confirm the CEF driven
$T_N$ behavior and the direction of ordered moments observed in other members of the series. We discuss also the
particular case where the rare-earth moments order out of the \textit{c}-axis, for which $T_N$ can be reduced by tuning
the CEF parameters. This fact represents an interesting frustration mechanism that may play some role in the
origin of the relevant low-dimensional SF in complex classes of materials such as the Ce$_{m}$MIn$_{3m+2}$ (M =
Rh, Ir and Co; \textit{m} = 1,2) HFS.

\section{\bf EXPERIMENT}
Single crystals of TbRhIn$_{5}$ with dimensions up to 2 cm$^{3}$ approximately were grown from In-flux, as
reported previously.\cite{Hegger,Petrovic1,Petrovic2} Most of the crystals show columnar habit, with their long
axis along the tetragonal \textit{c}-axis. The tetragonal HoCoGa$_{5}$-type structure \cite{Moshopoulou} with cell
parameters \textit{a} = 4.595(4) \r{A} and \textit{c} = 7.418(4) \r{A} were confirmed by x-ray powder diffraction and the
crystal orientation was determined by the usual Laue method. Specific heat measurements were performed in a
Quantum Design PPMS small-mass calorimeter that employs a quasi-adiabatic thermal relaxation technique.
Magnetization measurements were made in a Quantum Design \textit{dc} Superconducting Quantum Interference Device and
electrical resistivity was measured using the PPMS low-frequency \textit{ac} resistance bridge and four-contact
configuration. The samples used in the electrical resistivity measurements were previously screened for In-free
surface contamination.

RXMD measurements were carried out at the bending magnet beamline XRD2 of the Brazilian Synchrotron Light Source
(LNLS), in Campinas, Brazil using a double-bounce Si (111) monochromator, with sagittal focusing.\cite{Giles} A Rh-coated
mirror was used to vertically focus the beam and also to eliminate third and higher-order harmonics in the
incident beam. The bending magnet source delivers photon beams with a flux of $\sim{3}$x10$^{10}$ photons/s
around 8 keV at 100 mA in  spot of $\sim{0.6}$ mm (vertical) x 2.0 mm (horizontal) at the sample, with an energy
resolution from the source of $\frac{\delta E}{E}$ $\sim{10^{-4}}$. Our experiments were performed in the
vertical scattering plane, i.e, perpendicular to the linear polarization of the incident photons ($\sigma$ polarization\cite{Hill}).

Although neutron diffraction is the natural choice of experiments to resolve magnetic structures, RXMD offers
the advantage that only very small samples are required. For the RXMD experiments the high resolution of the
magnetic reflections is obtained as a natural consequence of the intrinsic collimation of the synchrotron x-ray
source, and the small cross section ensures that even for strong magnetic Bragg peaks, the intensity is not
extinction limited and a reliable measurement of the order parameter is possible. In our experiments high count
rates were obtained, allowing a precise temperature dependence of the ordered Tb moments.

A platelet of TbRhIn$_{5}$, from the same batch as that used for macroscopic properties measurements, was
mechanically polished perpendicular to \textit{c}-axis ((0 0 l) flat surface) to eliminate surface contamination from
the residual flux and to increase the reflectivity, which gives a mosaic spread characterized by the full width
at half maximum (FWHM) of $\sim{0.04}$°. The sample was cut parallel to the \textit{ab}-plane to have a final shape
of a long block with dimensions of 4x3x1 mm$^{3}$ and to investigate reflections in the [\textit{h} 0 \textit{l}] zone axis. The size of the sample was chosen to ensure that the beam completely reaches the crystal in the scattering plane for all angles of interest. The sample was mounted on the cold finger of a closed-cycle He cryostat (base
temperature 11 K) with a cylindrical Be window. 

\section{\bf RESULTS}

\begin{figure}[ht]
    \centering
        \includegraphics[width=0.45 \textwidth]{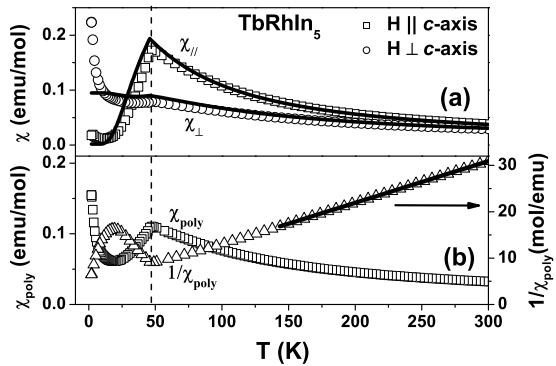}
  \caption{(a) Temperature dependence of the magnetic susceptibility for TbRhIn$_{5}$, for \textit{H} =1  kOe
applied parallel ($\chi_{//}$) (open squares) and perpendicular to the \textit{c}-axis ($\chi_{\perp}$) (open circles). The solid
lines are best fits to the data for both directions using our mean field model (see below). (b) $\chi_{poly}(T)$ and the
inverse 1/$\chi_{poly}(T)$; The solid line is the linear fit to the 1/$\chi_{poly}$ data for \textit{T}$
>$140 K. From this fit we extracted $\mu_{eff}$ = 9.4(1)$\mu_{B}$ and $\theta$ = - 47(1) K for TbRhIn$_{5}$.}
\label{fig:Fig1ChiEPS}
\end{figure}

The temperature dependence of the magnetic susceptibility measured for a magnetic field \textit{H} = 1 kOe applied
parallel $\chi_{//}$ and perpendicular $\chi_{\perp}$ to the \textit{c}-axis is presented in
Fig.~\ref{fig:Fig1ChiEPS}(a). Fig.~\ref{fig:Fig1ChiEPS}(b) shows the polycrystalline average of
Fig.~\ref{fig:Fig1ChiEPS}(a) data taken as $\chi_{poly}$ = ($\chi_{//}$ + 2 $\chi_{\perp}$)/3.

From a linear fit to the inverse of $\chi_{poly}$(\textit{T}) for \textit{T}$>$140 K using a Curie-Weiss law, we have
obtained a Curie-Weiss temperature $\theta$ = - 47(1) K and the Tb$^{3+}$ effective magnetic moment $\mu_{eff}$
= 9.4(1) $\mu_{B}$. As it can be seen in Fig.~\ref{fig:Fig1ChiEPS}(a) the magnetic susceptibility of
TbRhIn$_{5}$ is higher for the field applied along the \textit{c}-axis, in agreement to what was found for all others
non-S R-members of these series.\cite{Hegger,pagliuso3,pagliuso4} The ratio $\chi_{//}$/$\chi_{\perp}$ $\sim$
2.08 taken at $T_N$ is mainly determined by the tetragonal CEF and it reflects the same order of magnetic
anisotropy found for other members of these series.\cite{Hegger,pagliuso3,pagliuso4} The solid lines in
Fig.~\ref{fig:Fig1ChiEPS}(a) are the best fits to the data using a mean field model which
includes an isotropic first-neighbors exchange interaction and the tetragonal CEF.\cite{pagliuso5} The best fit
yields a J$_{R-R}$ = 0.2 meV (J$_{R-R}$ is equal to $zK_{RKKY}$ in the notation of Ref. \onlinecite{pagliuso5})
for the exchange parameter and the CEF parameters $B_{20}$ = -1.4 x 10$^{-1}$ meV, $B_{40}$ = 1.3 x 10$^{-4}$
meV, $B_{44}$ = -5.3 x 10$^{-3}$ meV, $B_{60}$ = 0.21 x 10$^{-4}$ meV, $B_{64}$ = 1.5 x 10$^{-4}$ meV. The CEF
level scheme obtained from the splitting of the Tb$^{3+}$ (\textit{J} = 6) multiplet by above parameters is built up of
three doublets and seven singlets with an overall splitting of roughly 310 K. The calculated curves using our
model reproduce very well the magnetic anisotropy and the peak of the magnetic susceptibility at $T_N$ $\sim$
45.5 K for both directions (Fig.~\ref{fig:Fig1ChiEPS}(a)). At lower-\textit{T} an intrinsic and anisotropic Curie-like
tail can be seen in the magnetic susceptibility data and it is presumably associated with an additional magnetic
transition with changes in the magnetic structure which happens within the ordered state (similarly to
TbIn$_3$ cubic compound at higher temperatures\cite{raimundo,galera}).

\begin{figure}[ht]
    \centering
        \includegraphics[width=0.45 \textwidth]{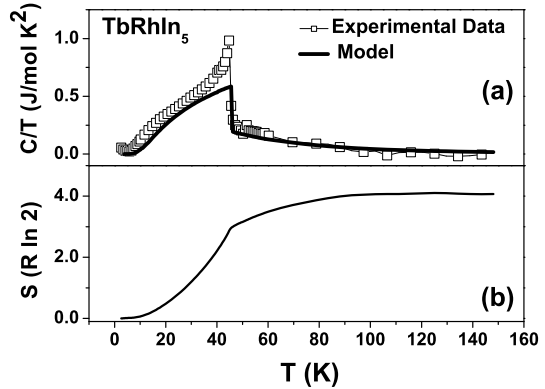}
    \caption{(a) Specific-heat data divided by temperature as a function of temperature for
    a single crystal of TbRhIn$_{5}$. The solid line is the best fit
 to this data using our mean field model. (b) The corresponding magnetic entropy in the temperature range
2$<$\textit{T}$<$150 K for TbRhIn$_{5}$.}
        \label{fig:Fig2HCEPS}
\end{figure}

Fig.~\ref{fig:Fig2HCEPS}(a) shows the specific heat divided by temperature and the corresponding magnetic
entropy (Fig.~\ref{fig:Fig2HCEPS}(b)) in the temperature range 0$<$\textit{T}$<$150 K for TbRhIn$_{5}$. To calculate
the magnetic entropy, the phonon contribution was estimated from the non-magnetic specific-heat data of
YRhIn$_{5}$ and subtracted from the total specific heat of the magnetic compound. The recovery magnetic entropy
at high-\textit{T} is close to its expected values for \textit{J} = 6. An anomaly in the specific-heat data associated with the onset of
AFM order can be seen at $T_N$= 45.5 K in good agreement with the temperatures where the maximum in the magnetic
susceptibility occurs (see Fig.~\ref{fig:Fig1ChiEPS}). Again, the solid line in Fig.~\ref{fig:Fig2HCEPS}(a)
represents the best fit to the data using our mean field model for the same parameters used in the fit of the
$\chi(T)$ data (the best set of parameters was obtained from simultaneous minimization process for both
$\chi(T)$ and \textit{C/T}(\textit{T}) data).

\begin{figure}[ht]
    \centering
        \includegraphics[width=0.45 \textwidth]{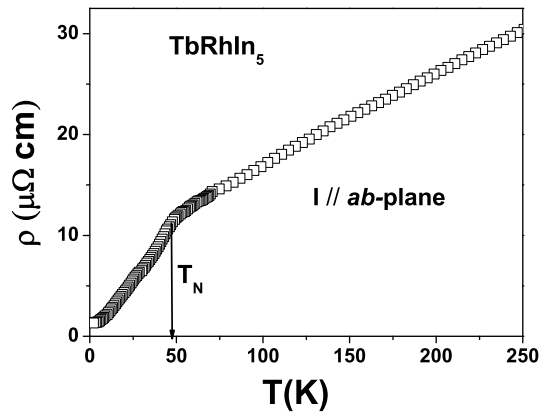}
    \caption{Temperature dependence of the electrical resistivity for TbRhIn$_{5}$ single crystal.
    The current (\textbf{I}) has been applied parallel to the \textit{ab}-plane. The solid arrow point out
    a kink at the N\'eel temperature for this compound.}
    \label{fig:Fig3RhoEPS}
\end{figure}

The temperature dependence of the electrical resistivity for TbRhIn$_{5}$ single crystals is plotted in
Fig.~\ref{fig:Fig3RhoEPS}. Among various measured crystals, the room temperature value of the resistivity ranges
between 10 -- 35 $\mu \Omega$ cm indicating the high quality of the crystals. At high temperature the data
always showed a metallic behavior while, at low temperatures, a clear kink can be seen at the ordering
temperature $T_N$.

\begin{figure}[ht]
    \centering
        \includegraphics[width=0.45 \textwidth]{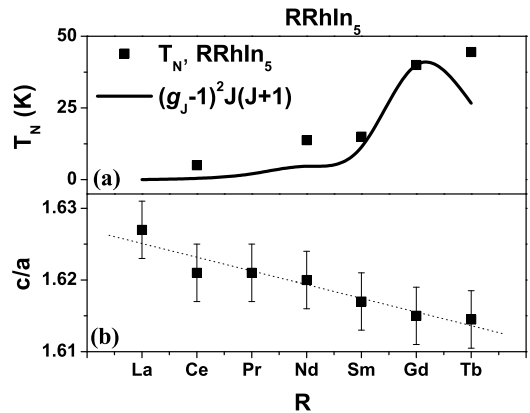}
    \caption{$T_N$(a) and \textit{c/a} ratio (b) for the RRhIn$_{5}$ compounds. The solid line in (a) represents the de Gennes
    factor $(g_{J}-1)^2J(J + 1)$ for the ground state multiplet \textit{J} of the rare earth ions, normalized
    by the GdRhIn$_{5}$. The dashed line in (b) is just a guide to the eye.}
    \label{fig:Fig4GennesEPS}
\end{figure}

The \textit{c/a} ratio and $T_N$ values for RRhIn$_{5}$ compounds are shown in Fig.~\ref{fig:Fig4GennesEPS}. The solid
line in Fig.~\ref{fig:Fig4GennesEPS}(a) is the expected behavior for $T_N$ and $\theta$ according to de
Gennes factor $(g_{J}-1)^2J(J + 1)$ for the ground-state multiplet \textit{J} of each rare earth normalized by their
values for GdRhIn$_{5}$. As for R = Ce and Nd,\cite{pagliuso4} $T_N$ of the TbRhIn$_{5}$ does not follow the de
Gennes scaling. Interestingly, $T_N$ for TbRhIn$_{5}$ is the highest among the existing RRhIn$_{5}$ members.

The microscopic low-temperature magnetism of TbRhIn$_5$ was further investigated by RXMD. For the sample orientation used in the experiments, with the zone axis [\textit{h} 0 \textit{l}] parallel to the scattering plane, the resonant scattering cross section (the non-resonant scattering contribution was observed to be negligible) for the case of a dipolar resonance with a linear polarized incident beam perpendicular to the scattering plane may be written as:\cite{Hill,Lovesey}
\begin{eqnarray} \label{eq:equation1}
 I\varpropto\arrowvert\sum_{n} {\hat{u}_{n}} \cdot {\hat{k}_{f}}e^{i{\bf{Q}} \cdot {\bf{r_n}}}\arrowvert^2
\end{eqnarray}
where $\bf{Q}=k_{f}-k_{i}$ is the scattering vector, $\hat{k}_{f}$ is the direction of the scattered wave vector
$\bf{k_{f}}$, and $\hat{u}_{n}$ is the moment direction at the \textit{n}th site. The proportionality symbol
$\varpropto$ includes \textbf{Q}-independent resonant amplitudes, the Lorentz factor, arbitrary scale factors and
an angular correction factor for asymmetric reflections.  The summation is over all the \textit{n}th resonant ions in
the magnetic unit cell and $\bf{r_n}$ is the position of such an ion. Note that in the present geometry and in the absence of an in-plane $\pi$ polarized component of the incident beam, the dipolar resonant cross section terms are not sensitive to the component of the ordered moment perpendicular to the scattering plane (i.e., along the \textit{b} axis).

Below $T_N$ $\sim$ 46 K, careful scans along [\textit{h} 0 \textit{l}] direction revealed superstructure Bragg peaks of type ($\frac{2n+1}{2}$ 0 $\frac{2m+1}{2}$) (\textit{n}, \textit{m} integers), appearing as a result of the strong resonant enhancements of the magnetic peaks for both Tb $L_{II}$ and $L_{III}$ edges, and the high magnetic moment per Tb$^{3+}$ ion of $\mu$=9.5 $\mu_{B}$. In the low temperature phase, magnetic peaks (0 $\frac{2n+1}{2}$ $\frac{2m+1}{2}$) were also observed, revealing a twinned magnetic structure. The intensities of ($\frac{2n+1}{2}$ 0 $\frac{2m+1}{2}$) and (0 $\frac{2n+1}{2}$ $\frac{2m+1}{2}$) were comparable, suggesting approximately equal domain population. Therefore, our magnetic cell is duplicated in the \textit{a} and \textit{c} directions when compared to the chemical one. Above $T_N$ only charge peaks consistent with the tetragonal HoCoGa$_{5}$-type structure were observed. The widths (FWHM) of magnetic peaks were the same as that for equivalent scans through a charge Bragg peak.

\begin{figure}[ht]
    \centering
        \includegraphics[width=0.45 \textwidth]{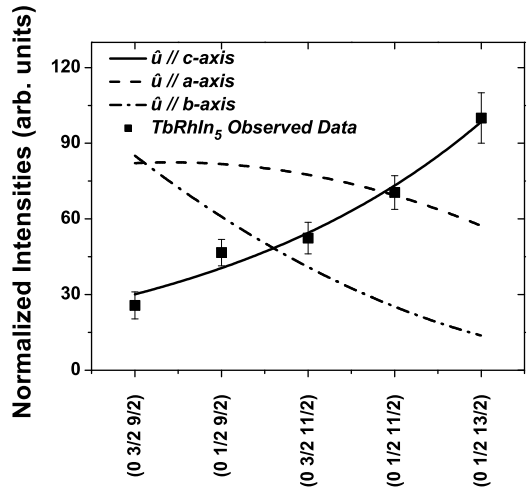}
    \caption{Integrated intensities of five experimentally-obtained magnetic reflections for TbRhIn$_{5}$
($\blacksquare$) along with calculated intensities for a magnetic moment parallel to \textit{c}-axis (solid line), to \textit{a} (dashed line) and \textit{b}-axis (dashed-dotted line).}
    \label{fig:Fig5IntensEPS}
\end{figure}

The magnetic structure of TbRhIn$_{5}$ was then resolved from the obtained modulation vector $\tau$=($\frac{1}{2}$ 0 $\frac{1}{2}$) and the magnetic moments direction was established by comparing the observed intensities of some magnetic Bragg reflections with a model based on the resonant cross section of Eq.~\ref{eq:equation1}. In Fig.~\ref{fig:Fig5IntensEPS}, the integrated intensities of five magnetic reflections obtained at \textit{T} = 20 K are shown with solid squares. We used asymmetric reflections (\textbf{Q} vector out of scattering plane) of the type (0 $\frac{2n+1}{2}$ $\frac{2m+1}{2}$) in order to also analyze the \textit{b}-axis component in the resonant cross section, which can not be distinguished with \textbf{Q} vectors into the scattering plane (i.e. with ($\frac{2n+1}{2}$ 0 $\frac{2m+1}{2}$) reflections). Experimental reflections were taken at the maximum enhancement of magnetic signal at the $L_{II}$ edge, \textit{E} = 8253 eV, and the data has been numerically integrated using gaussian fit functions (solid squares in Fig.~\ref{fig:Fig5IntensEPS}). Three possibilities were tested, a first one assuming the moments
along the \textit{c}-direction (solid line), a second with the moments aligned along the \textit{a}-direction (dashed line) and a third one with the moments along \textit{b}-direction (dashed-dotted line).
As can be seen in Fig.~\ref{fig:Fig5IntensEPS} the model with the ordered moments $\hat{u}//c$-direction is the
one in agreement with the observed data. A similar trend was also observed for NdRhIn$_{5}$ family
compound\cite{wei2} in its low-temperature AFM phase with the moment locked in the \textit{c}-axis. Given the
experimental errors, the directions of the ordered moments is determined within $\sim{10}$° of the
\textit{c}-direction.

\begin{figure}[ht]
    \centering
        \includegraphics[width=0.28 \textwidth]{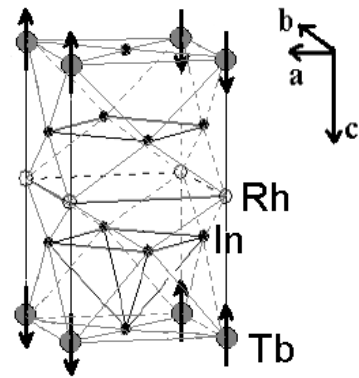}
    \caption{The magnetic structure of TbRhIn$_{5}$ at 20 K. A half magnetic unit cell is shown.
 $\uparrow$($\downarrow$) represents the directions ``up'' (``down'') of magnetic moments of Tb
 (depicted by gray filled circles) along the \textit{c} axis. }
    \label{fig:Fig6EstMagEPS}
\end{figure}

In Fig.~\ref{fig:Fig6EstMagEPS} the magnetic unit cell of TbRhIn$_{5}$ is shown according to results depicted in
Fig.~\ref{fig:Fig5IntensEPS}. A half magnetic cell is shown with moment directions at the Tb$^{3+}$ ion
crystallographic sites.

\begin{figure}[ht]
    \centering
        \includegraphics[width=0.45 \textwidth]{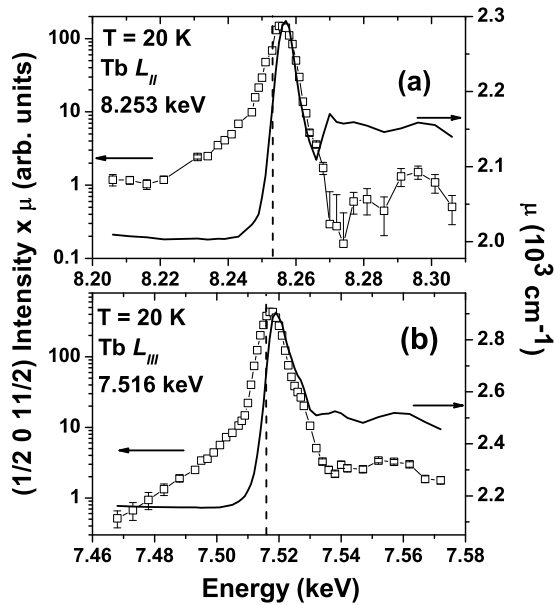}
    \caption{Energy dependence of the ($\frac{1}{2}$ 0 $\frac{11}{2}$) magnetic peak (open squares)
    through the Tb $L_{II}$ (a) and $L_{III}$ (b) edges. The open symbols are gaussian fit to the
elastic peak at each energy used in the scan. The data have been corrected for absorption. Solid line curve represents $\mu(E)$ obtained from the fluorescent yield (scale on the right). From the $\mu(E)$ inflection points we determine the absorption edges values (vertical dashed line) as being 8253 eV ($L_{II}$) and 7516 eV ($L_{III}$).}
    \label{fig:Fig7EnergEPS}
\end{figure}

In order to use the resonant enhancement of the magnetic peaks, the primary beam energy was tuned to the
$L_{II}$ and $L_{III}$ absorption edges of Tb$^{3+}$ ion (tabulated as being 8252 eV and 7514 eV, respectively). In
Figs.~\ref{fig:Fig7EnergEPS}(a) and (b) we plot the scattered intensity of the ($\frac{1}{2}$ 0 $\frac{11}{2}$)
magnetic Bragg peak as a function of the incident photon energy on tuning through the $L_{II}$ and $L_{III}$
edges, respectively. The data were corrected for absorption (coefficient $\mu(E)$ is showed with the solid line curve
and the right side scale in Figs.~\ref{fig:Fig7EnergEPS}) using the fluorescence emission.\cite{Evans}
The inflection points of $\mu(E)$ curves (vertical dashed line) were used to define the absorption edges. In both cases,
the maximum resonant enhancement is observed $\sim{2}$ eV above the edges, which is a signature of the electric
dipole (\textit{E}1) resonance involving electronic transitions 2$p_{1/2}\leftrightarrow$5$d$ and
2$p_{3/2}\leftrightarrow$5$d$. A remarkable enhancement of over two orders of magnitude for both edges has been
obtained. From the fit to a Lorentzian-squared line profile, the width of resonance were found
at both edges as being 6.2 eV and 6.4 eV for $L_{II}$ and $L_{III}$, respectively (corrected for energy
resolution).

The relative strength $I(L_{III})/I(L_{II})$ of the resonant enhancements (the ``branching ratio'') of the \textit{E}1
resonances is $\sim{2.9}$ which is in agreement with the theoretical expectations for branching ratios at the
rare-earth \textit{L} edges. This ratio is of interest since it relates to the underlying electronic and magnetic
structures (see Refs. \onlinecite{Veenendaal} and \onlinecite{McMorrow} and the references cited therein for
more details). According to Ref. \onlinecite{Veenendaal}, in heavy rare earth such as Tb, the \textit{E}1 resonance at
the $L_{III}$ is $\sim{2.5}$ times that of the $L_{II}$ edge. This expectation is in agreement with our
observations. Intensity oscillations have been observed above the
edges, and can be ascribed to magnetic DAFS oscillations\cite{Stragier} (Diffraction Anomalous
Fine Structure) which are oscillations of the anomalous scattering factors associated with the interference of
the photoelectrons wave function with the surrounding atoms that bring a fine structure of oscillations in the energy line-shape spectrum.

\begin{figure}[ht]
    \centering
        \includegraphics[width=0.45 \textwidth]{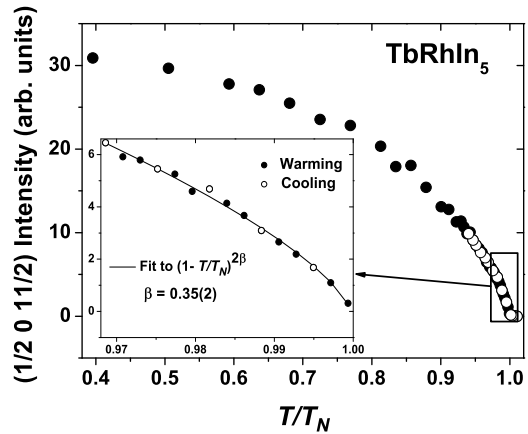}
    \caption{Temperature dependence of the magnetic Bragg integrated intensities for ($\frac{1}{2}$ 0
$\frac{11}{2}$) magnetic peak. The inset shows in details the critical region around the transition at the
N\'eel temperature for this compound. From a power law fit to this data within the temperature range of approximately 3\% below $T_N$ we obtain a $T_N$ = 45.56(2) K and a critical parameter $\beta$ = 0.35(2)}
    \label{fig:Fig8TempEPS}
\end{figure}

The behavior of the ordered phase was measured as the temperature was raised at the $L_{II}$ edge. The measurements
of the integrated intensities were performed on the ($\frac{1}{2}$ 0 $\frac{11}{2}$) satellite reflection. In
Fig.~\ref{fig:Fig8TempEPS} the data of longitudinal scans along this reflection is displayed as function of the
reduced temperature \textit{T}/$T_N$. As the temperature increases, the peak intensity gradually decreases and disappears
above $T_N$. This result, as well as the one obtained for the energy dependence described above,
clearly confirm the magnetic nature of the observed peaks. Data were taken on two regimes: on warming (filled
circles on Fig.~\ref{fig:Fig8TempEPS}) and cooling (open circles). A fitting to the usual power-law expression
($\sim(1-T/T_{N})^{2\beta}$) for a second order phase transition (denoted by a solid line in the inset) within the temperature range of approximately 3\% below $T_N$ on the warming regime data gives a magnetic transition temperature $T_N$ = 45.56(2) K and a critical order exponent $\beta$ = 0.35(2) for the Tb sublattice magnetization. This value of $\beta$ is compatible with a
three-dimensional Heisenberg system. No hysteresis was observed around the ordering temperature and the smooth
decreasing in intensity in the crossover to the paramagnetic phase is consistent with a second order transition.

In Table 1 we summarize the experimental parameters obtained for TbRhIn$_{5}$ in this work.

\section{\bf DISCUSSION}

As discussed in the introduction, to follow the evolution of the 4\textit{f} magnetism along the R$_{m}$MIn$_{3m+2}$
series is a crucial first step to achieve a deeper understanding of the complex physical properties of these
materials. Early comparative studies of the magnetic properties in this family have shown that for the Ce-based
materials where the magnetic ordered moments are not aligned along the \textit{c}-axis,\cite{wei2} $T_N$ is suppressed
to less than 0.5 of the CeIn$_{3}$ value for CeRhIn$_{5}$  where the Ce magnetic moments lie in the \textit{ab}-plane
within the AFM state. On the other hand, for NdRhIn$_{5}$, where the ordered moments point along the \textit{c}-axis
materials,\cite{wei2} $T_N$ is significantly enhanced when compared to that for their cubic NdIn$_{3}$ parent
compound. And lastly, for the Gd-based materials, where the CEF effects are small, the low temperature magnetic
properties remain nearly unaltered compared to the GdIn$_{3}$,\cite{pagliuso2,pagliuso4,granado2} suggesting
that the low temperature CEF scheme configuration plays a fundamental role in the observed trends.

\begin{table*}[t]
\caption{Experimental parameters for TbRhIn$_{5}$.} \label{table1}
\begin{ruledtabular}
\begin{tabular}{ccccccccccc}
 \textit{a}(\AA) & \textit{c}(\AA) & $T_N$(K) & J$_{R-R}$(meV) & $\frac{T_{N,RRhIn_{5}}}{T_{N,RIn_{3}}}$ & $B_{20}$(meV) & $B_{40}$(meV) & $B_{44}$(meV) & $B_{60}$(meV) & $B_{64}$(meV)
\\ \hline 4.595(3) & 7.418(3) & 45.5 & 0.2 & 1.39 & -1.4x10$^{-1}$ & 1.3x10$^{-4}$ & -5.3x10$^{-3}$ & 0.2x10$^{-4}$ & 1.5x10$^{-4}$
\end{tabular}
\end{ruledtabular}
\end{table*}

Regarding the influence of a given CEF scheme in the AFM ordering temperature, it is reasonable to assume that
if a system orders in a given direction and CEF parameters are modified making it more magnetically susceptible
in some other direction, but without actually changing the order, the system may experiment some kind of
magnetic frustration or the energy barrier between these states should diminish. Therefore, the ordering
temperature should decrease as well. Inversely, if the system orders in a certain direction and CEF parameters
change favoring even more this state, the ordering temperature should increase.

We have shown that this new member of the RRhIn$_{5}$, TbRhIn$_{5}$, orders antiferromagnetically at $T_N$
= 45.5 K, which is an enhanced ordering temperature when compared to $T_N$ $\sim$ 32 K of TbIn$_{3}$. As
TbRhIn$_{5}$ is more magnetically susceptible for a field applied along the \textit{c}-axis, according the the scenario
above, the Tb moment must be ordered along the \textit{c}-axis to explain the enhanced $T_N$ of TbRhIn$_{5}$.

Taking advantages of the higher \textbf{Q} resolution available with x-ray scattering techniques and using the
RXMD cross section to determine moment directions in magnetic compounds,\cite{Hannon,Hill} we resolved the
magnetic structure of TbRhIn$_{5}$. Recently, this technique has been also successfully used to
determine the magnetic structure of two Gd members of the R$_{m}$MIn$_{3m+2}$ (\textit{m} = 1, 2; M = Rh, Ir; \textit{n} = 1)
series.\cite{granado1,granado2}

As expected from the idea above, the solved magnetic structure of TbRhIn$_{5}$ reveals a commensurate
AFM structure with propagation vector ($\frac{1}{2}$ 0 $\frac{1}{2}$) and the Tb moments oriented
along the \textit{c}-axis (see Fig.~\ref{fig:Fig6EstMagEPS}). The direction of the ordered moments was established by comparing the observed intensities of five magnetic Bragg reflections (Fig.~\ref{fig:Fig5IntensEPS}) with a model based on the resonant cross section for the case of dipolar resonance with a linear polarized incident beam perpendicular to the scattering plane (Eq.~\ref{eq:equation1}). The model considers the possible orientation of the moments along the three tetragonal axis. In Fig.~\ref{fig:Fig5IntensEPS} it is obvious the agreement between experimental data and the model when the calculations were done assuming the moments along \textit{c}-direction (solid line).

In addition, a proposed mean field model including an isotropic first-neighbors exchange interaction
and the tetragonal CEF\cite{pagliuso5} has shown that the enhancement of $T_N$ for tetragonal compounds that
orders axially is a general trend for tetragonal materials when the CEF parameters tend to increase the fluctuations along the \textit{c}-axis. We have used this model to fit our susceptibility and specific heat data, and we could successfully reproduce all the main features of our data,
including the prediction of AFM ordering along the \textit{c}-axis\cite{pagliuso5} for the CEF and J$_{R-R}$ parameters
given in Table 1.

Therefore, the reported magnetic properties of TbRhIn$_{5}$ compound is another experimental evidence of
these general CEF induced trends along R$_{m}$MIn$_{3m+2}$ (\textit{m} = 1, 2; M = Rh, Ir). As such, with the CEF
effects being very important in determining $T_N$, it is expected that the de Gennes scaling would fail to
describe the behavior of $T_N$ along the RRhIn$_{5}$ series (see Fig.~\ref{fig:Fig4GennesEPS}).

Is is important to note that this mean field model has also predicted that, for \textit{J} = 5/2 and when the system
spin is on the plane (for the Ce case) the N\'eel temperature approximately decreases when the CEF parameters
increases fluctuations on \textit{c}-axis, which is an effective measure of the system likeness to be in the
\textit{c}-direction. This prediction is consistent with the fact that the tetragonal CeRhIn$_{5}$ and
Ce$_{2}$RhIn$_{8}$ present higher magnetic susceptibilities when magnetic field is applied along the \textit{c}-axis and
$T_N$ a factor of two smaller than that for CeIn$_{3}$. It is obvious that the hybridization and Kondo effects
are very important in the case of the Ce-based materials, but it is interesting to note that this CEF driven
magnetic frustration mechanism combined to hybridization could create strong in-plane magnetic fluctuations that
can mediate the quasi-2D unconventional superconductivity in these systems. It would be interesting to further
test this model for others members of the R$_{m}$MIn$_{3m+2}$ that present decreasing of $T_N$ for tetragonal
compounds. Although the effect is only about 20\%, the tetragonal Sm-based compounds present smaller $T_N$
values than their cubic relative SmIn$_{3}$.\cite{pagliuso4} Therefore, according to the present model, the
Sm-ordered moment for these materials should be aligned out of the \textit{c}-axis, most likely in the \textit{ab}-plane.

\section{\bf CONCLUSION}

In conclusion, we have presented the physical properties and magnetic structure of a new member of the
R$_{m}$MIn$_{3m+2}$ series, TbRhIn$_{5}$. This intermetallic compound orders antiferromagnetically at $T_N$
= 45.5 K, the highest ordering temperature among the existing RRhIn$_{5}$ materials. The solved magnetic
structure of TbRhIn$_{5}$ is commensurate with propagation vector $\tau$=($\frac{1}{2}$ 0 $\frac{1}{2}$) and the
Tb moments oriented along the \textit{c}-axis. The direction of the ordering Tb-moments and the enhanced $T_N$ of
this compound were successfully explained by a mean field model including an isotropic first-neighbors exchange
interaction and the tetragonal CEF. Also, the model reproduces all interesting features of our susceptibility and specific heat experimental data
for the set of CEF parameters given in Table 1. We argue that reported magnetic properties of TbRhIn$_{5}$ compound is another experimental evidence of a more general CEF induced trend along the R$_{m}$MIn$_{3m+2}$
(\textit{m} = 1, 2; M = Rh, Ir). The particular case where the rare-earth moments ordered out of the \textit{c}-axis and the
$T_N$ can be reduced by tuning the CEF parameters reveals a frustration mechanism that may play some role in
producing in-plane magnetic fluctuations relevant to the physical properties of complex classes of materials
such as the Ce$_{m}$MIn$_{3m+2}$ (M = Rh, Ir and Co; \textit{m}=1,2) HFS.

\begin{acknowledgments}
We thank A. O. G. Rodriguez and J. Madureira for helpful and interesting discussions about the magnetic moments
orientation model. This work was supported by FAPESP (SP-Brazil) Grants No.  03/09861-7, 04/08798-2, 05/00962-0, 
00/08649-6 and CNPq(Brazil) Grants No. 140613/2002-1, 307668/03, 04/08798-2 and 304466/20003-4. LNLS at Campinas is also acknowledged for beamtime at XRD2 beamline.

\end{acknowledgments}

\bibliography{Bibliog}

\end{document}